\documentstyle[aps,prb,multicol,graphicx]{revtex}

\draft

\begin{document}

\title{The Hubbard model applied to the phase diagram and pressure effects
in $YBa_2Cu_3O_{7-\delta}$ superconductors}

\author{E. S. Caixeiro and E. V. L. de Mello } 

\address{Departamento de F\'{\i}sica,
Universidade Federal Fluminense, av. Litor\^ania s/n, Niter\'oi, R.J.,
24210-340, Brazil}
 
%\thanks[caixa] {Partially supported by the Brazilian
%agencies Capes, Faperj and CNPq.}  
\date{Received \today }
\maketitle

\begin{abstract}
We apply a method based on a BCS-type approach to the extended Hubbard model
on a square lattice to deal with the $YBa_2Cu_3O_{7-\delta}$ family of
superconductors under pressure. The parameters of the tight-binding band are
taken from experiments, and the coupling strength $U$ and $V$ are estimated by
the zero pressure phase diagram ($T_c\times n_h$). This scheme yields the
nontrivial dependence of the superconductor critical temperature $T_c$ as a
function of the hole concentration $n_h$ in the $CuO_2$ plane. With the
assumption that the pressure $P$ modifies the potential $V$ and the on plane
hole content $n_h$, we can distinguish the charge transfer and the intrinsic
contribution to $T_c(P)$. We show that the changes on $T_c(P)$ for the
$YBa_2Cu_3O_7$ optimally doped compound at low pressures are almost entirely
due to the intrinsic term. \end{abstract}


\begin{multicols}{2}
\section{Introduction}
Pressure experiments have played an important hole in understanding high-$T_c$
superconductors (HTSC). Several review papers have been written on this
subject~\cite{Schilling,Takahashi,Wijngaarden}. In short, the pressure
investigation is a helpful tool to display the potential effects of the
chemical pressure, thereby providing useful information for synthesis of
related compounds and, more important to the present work, to understand which
are the structural parameters that influence the microscopic superconducting
mechanism responsible for the HTSC.  

It is well known that the pressure effects are twofold: (1) It is
documented that the applied pressure $P$ increases the hole
concentration $n_h$ on the $CuO_2$ planes~\cite{Wijngaarden}, with a maximum
rate of charge transfer approximately given by ${\partial n_h / \partial P
}=0.02{GPa}^{-1}$. This is called the pressure induced charge transfer 
(PICT)~\cite{Almasan} term. (2) Another pressure effect is the increase of $T_c$
for the optimum doped compound ($n_{op}$) above the zero pressure value, which
indicates another mechanism independent of the PICT, known as the
``intrinsic" contribution term~\cite{33}. Recently, a very large intrinsic term of
${\partial T_c / \partial P }=6.8K{GPa}^{-1}$ has been
measured~\cite{Orlando}, however, its origin is still a matter of research.
Usually these two effects, the PICT and the intrinsic term, have to be combined
in order to account for the experimental data on a given family of compounds.
It is very convenient to separate both contribution because the PICT can be
measured independently by thermal measurements, like the Hall coefficient.
Some earlier works have estimated the magnitude of both effects through a
phenomenological pressure expansion~\cite{33,34} applied to the
$YBa_2Cu_3O_{7-\delta}$ family of compounds. 

A method using a BCS-type mean field approach with an extended Hubbard
Hamiltonian was introduced to derive the phase diagram $T_c\times
n_h$~\cite{27,deMello2}. This procedure was used to deal with the pressure
effects and it was assumed that, besides the hole content $n_h$,
the pressure also changes the magnitude $V$ of
the attractive potential, and ultimately changes
the zero temperature superconducting gap $\Delta (0)$~\cite{32}. Thus, the
critical temperature $T_c(n_h,P)$ was obtained as an expansion in powers of
$P$, for a determined $n_h$ value, and each contribution was singled out,
which is very convenient to interpret the experimental results. This method
was applied to the experimental results of Hg 1201 and Hg 1223, with a fixed
set of parameters, with excellent agreement with the underdoped and overdoped
compounds~\cite{32}. In a similar approach, also using the BCS method and
the extended Hubbard Hamiltonian, Angilella et al~\cite{27} used the
isothermal compressibility tensor to estimate the change of $n_h$ with the
pressure and, through the derived $T_c\times n_h$, they obtained the variation
of the attractive potential $V$ with the pressure $P$. In their
calculations they have also taken into account pressure dependence for the
lattice parameter $a_i$, so that the hopping integrals $t_i$ depends on the
pressure too ($t_i=t_i(a_i(P))$). In this way, they could separate the
charge transfer and the intrinsic contribution. Their calculations for the
intrinsic term were in good agreement with the measurements of
the Bi 2212~\cite{Huang}. 

Calculations based on the extended Hubbard Hamiltonian provide a
possible microscopic interpretation for the intrinsic term since it is related
with the superconducting interaction $V$. Through such calculations, one may
extract some clues to the behavior of the fundamental microscopic mechanism
under the structural changes provided by the applied pressure, and some hints
for the mechanism itself.

Recently, we have combined both approaches of Ref.~\cite{27,deMello2} in order
to study the measured $T_c\times P$ curves of the
$Tl_{0.5}Pb_{0.5}Sr_2Ca_{1-x}Y_xCu_2O_7$ series~\cite{Edson}. A characteristic
of this method is that, if the phase diagram is experimentally known, $T_c(P)$
can be calculated for the entire family of different $n_h$ compounds, and the
value of $\partial n_h/\partial P$ calculated for this specific family. Our
calculated $T_c(n_h,P)$ for the $Tl_{0.5}Pb_{0.5}Sr_2Ca_{1-x}Y_xCu_2O_7$ 
series agreed well with the experiments for the extended-$s$ wave
symmetry~\cite{19}.

In a recently letter, Chen et al~\cite{Chen} have calculated the
pressure effects on the nearly optimum doped $YBa_2Cu_3O_7$ and underdoped
$YBa_2Cu_4O_8$, using a BCS-type method. Using values of $\partial
n_h/\partial P$ averaged from different results of several groups, they
obtained very good agreement with the experimental data. However, calculating 
$\partial T_c/\partial P$ ($P\approx 0$) for different $n_h$ compounds of the
$YBa_2Cu_3O_{7-\delta}$ family, they found almost the same values for
different potential parameters, which led them to conclude that the main
contribution to $T_c(n_h,P)$ was from the PICT term. To check whether the
$YBa_2Cu_3O_{7}$ compound has an intrinsic contribution term or not we have
performed calculations on the YBCO system using our method.

In our method we use a BCS approach with the extended Hubbard model to
calculate the phase diagram of the $YBa_2Cu_3O_{7-\delta}$ system, which is
compared with the experimental data. Then we take into account the effects
produced by the pressure through a change in the intersite potential: as
first order approximation we write $V(P)=V+\Delta V(P)$. Here $\Delta
V(P)=c_1P$, with $c_1$ being a constant independent of $P$ that will be
defined later. Since the structural changes for typical pressures are small,
we neglect its effect on the hopping integrals. For the density of hole
carriers we use the well know dependence of $n_h$ with $P$~\cite{34}. From
this we estimate the critical temperature $T_c(n_h(P),V(P))$ as an expansion
in terms of the pressure $P$. Our results showed good agreement with the
experimental data.

\section{The Phase Diagram } 

To develop the dynamics of the hole-type carriers in the Cu-O planes, we
adopt a two dimension extended Hubbard Hamiltonian~\cite{27,32,36} in a
square lattice of lattice parameter $a$

\begin{eqnarray} 
H&=&-\sum_{\ll ij\gg \sigma}t_{ij}c_{i\sigma}^\dag
c_{j\sigma}+U\sum_{i}n_{i\uparrow}n_{i\downarrow}
\nonumber \\
&& +\sum_{<ij>\sigma
\sigma^{\prime}}V_{ij}c_{i\sigma}^\dag c_{j\sigma^{\prime}}^\dag
c_{j\sigma^{\prime}}c_{i\sigma}, \label{b}    
\end{eqnarray} 
where $t_{ij}$ is the nearest-neighbor and next-nearest-neighbor hopping
integral between sites $i$ and $j$; $U$ is the Coulomb on-site correlated
repulsion and $V_{ij}$ is the attractive interaction between nearest-neighbor
sites $i$ and $j$. 

The transformation of Eq.(\ref{b}) to the momentum space leads to the
appearance of the dispersion relation. In order to compare with YBCO system,
we used a dispersion relation within a tight-binding approximation, which may
be compared with the one estimated from ARPES
measurements~\cite{Schabel}. Thus,  

\begin{eqnarray}
\varepsilon_{\bf k} &=& -2t_1(\cos(k_x a)+\cos(k_y a))+4t_2\cos(k_xa)\cos(k_ya)
\nonumber \\
&& -2t_3(\cos(2k_x a)+\cos(2k_y a))
\nonumber \\
&& +4t_4\cos(2k_xa)\cos(2k_ya)
\nonumber \\
&& -4t_5(\cos(2k_x a)\cos(k_y a)+
\nonumber \\
&& \cos(2k_ya)\cos(k_x a))-\mu ,
\label{e}
\end{eqnarray}
where it was considered identical hopping integrals along both directions
in the Cu-O planes for the nearest-neighbor ($t_x$=$t_y$=$t_1$), a different
one for the next-nearest-neighbor ($t_2$), and so on. Here it was considered a
hopping from the first to the fifth neighbor. Equation (\ref{e}) is in
agreement with the one given by Schabel et al~\cite{Schabel}; $\mu$ is the
chemical potential which controls the hole concentration.

Like the low temperature superconductors, the HTSC exhibit an energy ``gap"
(order parameter) in the superconducting phase which separates the paired
states from the single-particle states. Using a BCS-type mean-field
approximation~\cite{26} to develop Eq.(\ref{b}) in the momentum space, one
obtains the self-consistent gap equation, at finite temperatures~\cite{26}

\begin{equation}
\Delta_{\bf k}=-\sum_{\bf k^{\prime}}V_{\bf
kk^{\prime}}\frac{\Delta_{\bf k^{\prime}}}{2E_{\bf
k^{\prime}}}\tanh\frac{E_{\bf k^{\prime}}}{2k_BT},\label{cc}
\end{equation}
with
\begin{equation}
E_{\bf k}=\sqrt{\varepsilon_{\bf k}^2+\Delta_{\bf k}^2}, \label{rr}
\end{equation}
which contains the interaction potential $V_{\bf
kk^{\prime}}$ that comes from the transformation to the momentum space of
Eq.(\ref{b})~\cite{27,Schneider} 

\begin{equation}
V_{\bf kk^{\prime}}=U+2V\cos(k_xa)
\cos(k_x^{\prime}a)+2V\cos(k_ya) \cos(k_y^{\prime}a), \label{d}
\end{equation}

\begin{figure}
\includegraphics[width=8cm]{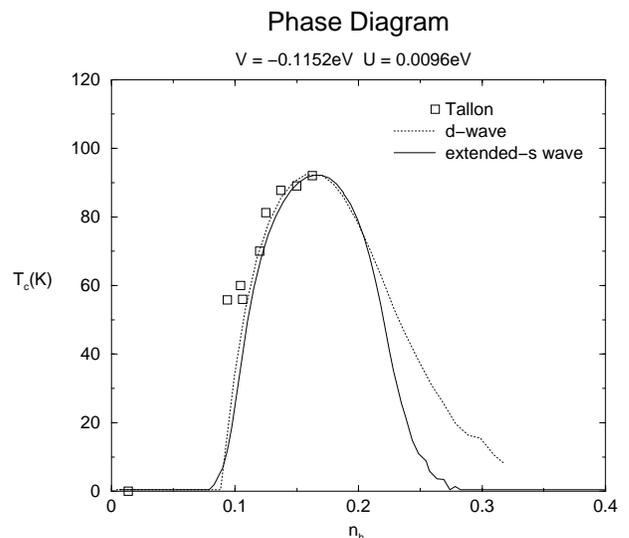}
\caption{Phase diagram $T_c \times n_h$ for the $d$ wave and
extended-$s$ wave symmetries, with the experimental data of the
$YBa_2Cu_3O_{7-\delta}$ taken from Ref.~\cite{Tallon} (open squares).} 
\end{figure}

The substitution of Eq.(\ref{d}) into Eq.(\ref{cc}) leads to appearance of a
gap with two distinct symmetries~\cite{27}:

\begin{equation}
\Delta_{\bf k}(\mu,T)=\Delta^{max}(\mu,T)[\cos (k_xa)\pm \cos (k_ya)],
\label{GG} \end{equation}
where the plus sign is for extended-$s$ wave and the minus sign, for
$d$ wave symmetry. In accordance with Ref~\cite{27} one observes that the $d$
wave part of the gap do not depend on the coupling constant $U$, depending only
on $V$. The extended-$s$ symmetry depends on both $U$ and $V$. As it is well
known~\cite{evandro,evandro2} at the critical temperature there is no symmetry
mixture, and the gap might be in the extended-$s$ or $d$ wave
state when $T\rightarrow T_c$.

Using the same BCS-type mean-field approximation used in the gap
equation one obtains the hole-content equation~\cite{24}

\begin{equation} 
n_h(\mu,T)=\frac{1}{2}\sum_{\bf k}\left(
{1-\frac{\varepsilon_{\bf k}}{E_{\bf k}} \tanh\frac{E_{\bf k}}
{2k_BT}}\right), \label{f}
\end{equation}
where $0\leq n_h\leq 1$. This equation, together with the gap equatios and the
dispersion relation, are solved numerically self-consistently, in the limit of
$T\rightarrow T_c$, in order to obtain the phase diagrams for both gap 
symmetry.

To compare with the experimental results for the Y123 system, we perform our
calculations with the hopping integrals: $t_1$=0.32eV as the
nearest-neighbor value; for $t_2$ it was adopted the ratio
$t_2/t_1$=0.50 for the $d$-wave symmetry and $t_2/t_1$=0.57 for the
extended-$s$; the remaining band parameters followed: $t_3/t_1$=0.16,
$t_4/t_1$=0.11, and $t_5/t_1$=0.031. All these values are close to the
ARPES measurements on Y123 system~\cite{Schabel}. 

Using these band parameters we first estimated the coupling constant $V$, using
the $d$-wave gap symmetry, reproducing the experimental phase diagram of
Ref.~\cite{Tallon}. The best value obtained was $V$=-0.1152eV. For the case of
the extended-$s$ gap symmetry, the value $U$=0.0096eV was obtained, using the
same coupling constant $V$ of the $d$-wave. The chemical potential $\mu$ is
calculated self-consistently. Fig.1 shows the results of our numerical
calculations together with the experimental data of Ref.~\cite{Tallon} for the
$YBa_2Cu_3O_{7-\delta}$ series. 

\section{ The Method }
To calculate the variations of $T_c$ for a given compound with a certain value
$n_h$, and under pressure $P$, we may use an expansion of $T_c(n_h,P)$ in
powers of $P$. Therefore,

\begin{eqnarray}
T_c(n_h,P)&=&T_c(n_h,0)+\left ( \frac{dT_c}{dP}\right
)_{P=0}P+
\nonumber \\ 
&&\frac{1}{2!}\left ( \frac{d^2T_c}{dP^2}\right )_{P=0}P^2
+\cdot \cdot \cdot,
\label{k} 
\end{eqnarray}
where

\begin{equation}
{d^zT_c\over dP^z}=\left ( c_1{\partial \over
\partial V}+ c_2{\partial \over \partial n_h}\right
)^zT_c(n_h(P),V(P)). \label{l} 
\end{equation}
Here, $c_1=({\partial V/\partial P})$ and $c_2={(\partial n_h/\partial
P)}$ are constant parameters, which are determined fitting the experimental
data for a given system. $T_c(n_h,0)$ is the critical temperature for $P$=0.
Equation (\ref{k}) can be written in a compact form, as  

\begin{equation}
T_c(n_h,P)=\sum_z\alpha_z{P^z\over z!}, \label{m} 
\end{equation}
with

\begin{equation}
\alpha_z=\left ( c_1{\partial \over \partial
V}+c_2{\partial \over \partial n_h}\right )^zT_c(n_h(P),V(P)). \label{n} 
\end{equation}

The first coefficient ($z=1$) for the above expression is given by 

\begin{equation}
\alpha_1=\left ( c_1{\partial T_c\over \partial V}+c_2{\partial T_c\over
\partial n_h}\right ). \label{p}  
\end{equation}
It is important to stress that the first term in $\alpha_1$ is the intrinsic
contribution, and the second one is the PICT contribution. Therefore, both
contribution can be singled out. Restricting ourselves to small changes $V$,
we may approximate  

\begin{equation}
{\partial T_c\over \partial V}\approx \overline{\Delta T_c\over \Delta V},
\label{q} 
\end{equation}

\begin{figure}
\includegraphics[width=8cm]{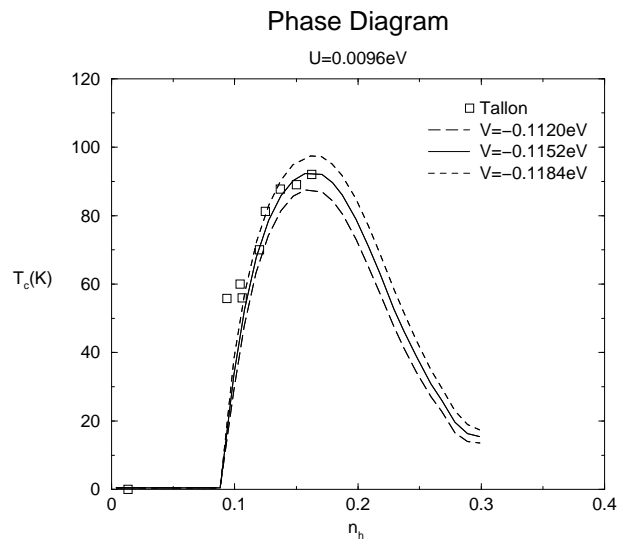}
\caption{Phase diagram $T_c \times n_h$ for the $d$-wave showing the effect of
a change on the coupling constant $V$, together with the experimental data of
the $YBa_2Cu_3O_{7-\delta}$ taken from Ref.~\cite{Tallon} (open squares).} 
\end{figure} 
where the horizontal bar denotes a ``mean" over the values of
$T_c$ obtained direct from the phase diagram as a function of $V$, as it is
shown in Fig.2. This ``mean" is realized for each value of $n_h$. To obtain
$\partial T_c/\partial n_h$ we may either use an experimental value, or use
directly the curves of Fig.2, as done here. One can also use a phenomenological
universal parabolic relation between $T_c$ and $n_h$~\cite{Edson,19}. For
the coefficient $\alpha_2$ we have

\begin{equation}
\alpha_2=\left ( c_1{\partial \over \partial V}+c_2{\partial \over \partial n_h}\right ) 
\left ( c_1{\partial T_c \over \partial V}+c_2{\partial T_c \over \partial
n_h}\right ). \label{u}
\end{equation}
Thus, $\alpha_2$ becomes 

\begin{equation}
\alpha_2\approx c_2^2{\partial^2 T_c \over \partial n_h^2}. \label{u}
\end{equation}
The other coefficients of the expansion can be obtained using the same
procedure.

\section{ The Pressure Experimental Data}
First of all, one observes that, at low pressures only the linear term
($\alpha_1$) of the expansion $T_c(n_h,P)$ comes into play. Therefore, we can
approximate $\alpha_1$ as the slope of the initial points of the $T_c\times
P$ curve, for a given $n_h$ compound. Starting with $n_h$=$n_{op}$, we
determine $c_1$ using the estimated $\alpha_1$ in Eq.(\ref{p}), as long as at
$n_{op}$ the charge transfer term vanishes. To determine $c_2$
we obtain $\alpha_1$ from the $T_c\times P$ curve of an $n_h\not= n_{op}$
compound, and use again Eq.(\ref{p}). Once these two constants are determined,
the $\alpha_z$ coefficients for any $n_h$ value can be calculated
(Eq.(\ref{n})). 

For the $YBa_2Cu_3O_{7-\delta}$ system we only have the non zero pressure data
for the $n_h$=$n_{op}$=0.165 ($\delta \approx 0$) and $n_h$=0.15 ($\delta
\approx 0.15$) compounds~\cite{Klotz,Tissen}. Thus, from the $T_c\times P$
curves of Ref.~\cite{Klotz,Tissen} we estimate $\alpha_1 = 0.75 K/GPa$ and
1.4 $K/ GPa$ for $n_{op}$ and $n_h$=0.15, respectively. The phase diagram
parameters ($\overline{\Delta T_c\over \Delta V}$) taken from Fig.2 and used
in the calculations were 1652$K/eV$ and 1475$K/ eV$ for the $d$-wave
and 1164$K/eV$ and 1099$K/eV$ for the extended-$s$ wave for $n_{op}$ and
$n_h$=0.165 compounds, respectively. Therefore, the resulting constants for
the $d$-wave were $c_1=4.54\times10^{-4}eV/GPa$ which furnishes an intrinsic
contribution of 0.8$K/ GPa$, and $c_2=3.15\times10^{-3}GPa^{-1}$ which is the
charge transfer term ($c_2={\partial n_h\over \partial P}$). For the
extended-$s$ we obtained $c_1=6.4\times10^{-4}eV/GPa$ and
$c_2=2.2\times10^{-3}GPa^{-1}$. It is important to mention that our values for
$\partial T_c^i/ \partial P$ (the intrinsic contribution; see Eq.(\ref{p}))
were in the same order of magnitude of Neumeier and Zimmermann~\cite{33},
despite the difference in the approaches.

\begin{figure} 
\includegraphics[width=8cm]{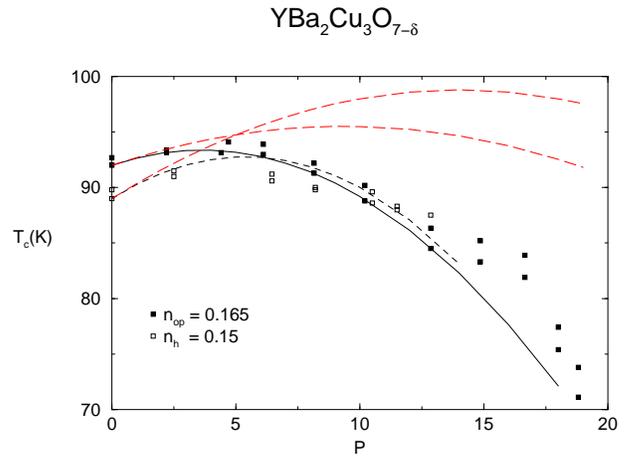}
\caption{Numerical results: the solid and dashed lines are the $d$-wave
results, while the long-dashed lines are the extended-$s$ wave
results. The open squares and filled squares are experimental data
of $YBa_2Cu_3O_{7-\delta}$ taken from Ref.~\cite{Klotz} and Ref.~\cite{Tissen},
respectively.}  
\end{figure} 

On Fig.3 we compare the $T_c(n_h,P)$ expansion with the
experimental data. We observe a very good agreement for both compounds,
specially for $n_{op}$, using the $d$-wave gap symmetry, where the data can be
fitted up to 20 $GPa$. The increase of $n_{op}$ with the pressure is attributed
to the intrinsic contribution in our theory, and the maximum $T_c$ at $P\approx
7GPa$ and subsequent decrease is due to the negative second order term of the
expansion, which reflects the effect of the PICT. The extended-$s$ wave
results did not show good results indicating a preference for the $d$-wave
pairing mechanism for the Y-123 system.   

\section{Conclusions}

We conclude this letter emphasizing that our method based on the BCS
approach with the extended Hubbard model, with both the attractive potential
$V$ and the density of carriers $n_h$ depending on the pressure $P$,
predicted an intrinsic term and a PICT contribution similar to those predicted
earlier ~\cite{33,34}. In our method we propose a direct pressure dependence
for the attractive potential $V$, which is different from other
works~\cite{27} that use $V$ as an adjustable parameter in order to fit the
known experimental dependence of $T_c$ on $P$. Our results for the $d$-wave are
in excellent agreement with the Y-123 pressure data and, the intrinsic term
being the most important contribution to the low pressure data on the
optimum compound. Concerning the underdoped compound the intrinsic
and PICT contribution are both present and must, as previously known by several
experiments and theories~\cite{Wijngaarden,33}, be taken into account. Such
results support the $d$-wave symmetry in these compounds.

We thank CNPq, Capes and FAPERJ for finacial support.

\end{multicols}
\end{document}